\documentclass[aps,prx,twocolumn,showpacs,showkeys,preprintnumbers,superscriptaddress,longbibliography]{revtex4}

\usepackage{graphicx,graphics,epsfig,subfigure,times,bm,bbm,amssymb,amsmath,amsfonts,amsthm,mathrsfs,MnSymbol}

\usepackage{times}
\usepackage{caption}
\usepackage{comment}

\newcommand{\be}{\begin{equation}}
\newcommand{\ee}{\end{equation}}
\newcommand{\ba}{\begin{eqnarray}}
\newcommand{\ea}{\end{eqnarray}}
\newcommand\tr{{\mbox{Tr\,}}}
\newcommand{\ignore}[1]{}
\newcommand{\bea}{\begin{eqnarray}}
\newcommand{\eea}{\end{eqnarray}}

\def\ii{\mathrm{i}}

\usepackage[pdfstartview=FitH]{hyperref}
\usepackage{pifont}
\hypersetup{
    colorlinks=true,       	
    linkcolor=red,          	
   citecolor=magenta,        
    filecolor=magenta,      	
    urlcolor=cyan,           	
    runcolor=cyan
}
\usepackage[pdftex]{color}
\usepackage{braket}
\usepackage{enumerate}

\newcommand{\blue}[1]{{\color{blue} #1}}

\newcommand{\eu}{{\rm e}}
\newcommand{\de}{{\displaystyle\rm\mathstrut d}}

\begin{document}

\title{Local convertibility and the quantum simulation of edge states in many-body systems}

\preprint{MIT-CTP 4463}

\author{Fabio Franchini}
\email{fabiof@mit.edu}
\affiliation{Department of Physics, Massachusetts Institute of Technology, Cambridge, MA 02139, U.S.A.}
\affiliation{SISSA and I.N.F.N, Via Bonomea 265, 34136, Trieste, Italy}
\author{Jian Cui}
\affiliation{Institute of Physics, Chinese Academy of Sciences, Beijing 100190, China}
\affiliation{Freiburg Institute for Advanced Studies, Albert Ludwigs University of Freiburg, Albertstra{\ss}e 19, 79104 Freiburg, Germany}
\author{Luigi Amico}
\affiliation{CNR-MATIS-IMM \& Dipartimento di Fisica e Astronomia, Via S. Soa 64, 95127 Catania, Italy}
\affiliation{Centre for Quantum Technologies, National University of Singapore, 3 Science Drive 2, Singapore 117543}
\author{Heng Fan}
\email{hfan@iphy.ac.cn}
\affiliation{Institute of Physics, Chinese Academy of Sciences, Beijing 100190, China}
\author{Mile Gu}
\affiliation{Center for Quantum Information, Institute for Interdisciplinary Information Sciences, Tsinghua University, Beijing 100084, China}
\affiliation{Centre for Quantum Technologies, National University of Singapore, 3 Science Drive 2, Singapore 117543}
\author{Vladimir  Korepin}
\email{korepin@gmail.com}
\affiliation{C. N. Yang Institute for Theoretical Physics, Stony Brook University, NY 11794, USA}
\author{Leong Chuan Kwek}
\affiliation{Centre for Quantum Technologies, National University of Singapore, 3 Science Drive 2, Singapore 117543}
\affiliation{National Institute of Education and Institute of Advanced Studies,
Nanyang Technological University, 1 Nanyang Walk, Singapore 637616}
\author{Vlatko Vedral}
\affiliation{Atomic and Laser Physics, Clarendon Laboratory, University of Oxford, Parks Road, Oxford, OX13PU, United Kingdom}
\affiliation{Centre for Quantum Technologies, National University of Singapore, 3 Science Drive 2, Singapore 117543}

\pacs{03.67.Mn, 03.67.Lx, 75.10.Pq, 03.67.Ac}

\begin{abstract}
In some many-body systems, certain ground state entanglement (R\'{e}nyi) entropies increase even as the correlation length decreases. This entanglement non-monotonicity {is a potential indicator of non-classicality}.
In this work we demonstrate that such a phenomenon, known as non-local convertibility, is due to the edge state (de)construction occurring in the system.
To this end, we employ the example of the Ising chain, displaying an order-disorder quantum phase transitions.
Employing  both analytical and numerical methods, we compute entanglement entropies for various  system bipartitions $(A|B)$ and consider ground states with and without Majorana edge states.
We find that  the thermal ground states, enjoying the Hamiltonian symmetries, show non-local convertibility if either $A$ or $B$ are smaller than, or of the order of, the correlation length. In contrast, the ordered (symmetry breaking) ground state is always locally convertible.
The edge states behavior explains all these results and could disclose a paradigm to understand local convertibility in other quantum phases of matter.
The connection we establish between convertibility and non-local, quantum correlations
provides a clear criterion of which features a universal quantum simulator should possess to outperform a classical machine.
\end{abstract}

\keywords{Local Convertibility | Universal Simulator | Edge States | Majorana | Ising Model | Entanglement }

\maketitle

In 1982, Richard Feynman conjectured that a quantum machine is necessary to predict the outcome of a general quantum evolution and pioneered the notion of a universal quantum simulator: a device capable of processing quantum information that potentially supersedes any classical computer in simulating quantum systems. This idea embraces much of quantum information research and the technologies stemming from it \cite{feynman1982,lloyd1996} and has attracted a lot of efforts toward the realization of such a device.
However, quantifying to what extent a given quantum system could outperform a classical simulator is problematic \cite{algorithm}.
How can we determine if a many body system can operate as an efficient quantum simulator?
To what extent is coherent manipulation the defining property of a quantum algorithm?
We address such a question quantitatively, using the local convertibility of the quantum system hosting the simulation, and we demonstrate that  the  (Majorana) edge states establishes genuinely quantum long-range correlations, that may provide an additional resource for a given computational protocol.

While quantum complexity and quantum algorithms theory has provided very general results on the computational power of abstract models\cite{Qcomplexity, Kitaev_QMA, Aharonov} , specific toy models (often providing concrete and physically relevant physical models) afford instances which exhibit interesting behavior in their own right (see f.i. Ref.\onlinecite{molphysics2011,chemistrysim}).
In the present work, we follow the later avenue.
Notwithstanding quantum many body systems provide a natural setting for  entanglement and other quantum superposition/interference effects \cite{entanglement1,entanglement2,iontraps,coldatoms,superconductivity,photonics,nmr},
it is poorly understood which one of the quantum resources  can indeed play as the added value for the simulation
\cite{algorithm,vidal2003a,vandernest2007,Briegel08,winter,eisert,Briegel09,vandernest2013,Raussendorf13,gottesman97}.
On the other hand, an important achievement has been the identification of the role of short- and long-range entanglement, which have also emerged as figures of merit for the different quantum orders that can be established in extended systems \cite{Wen09}.

{Here, we} refer to a specific notion of long range entanglement related to Local Operations and Classical Communications (LOCC) \cite{NielsenChuang}.
Upon partitioning a {many-body spin system} into two blocks $A$ and $B$, we consider the question: can the response of the ground state $|g\rangle$ to an external perturbation be rendered through LOCC restricted to $A$ and $B$ individually? If affirmative, the ground state can be moved around within a given quantum phase by LOCC. If negative, the adiabatic evolution induced by the perturbation involves some coherent quantum operation between system A and system B.
The figure of merit for such phenomena is the  differential local convertibility (DLC) of bipartite states. {DLC} was  introduced in the context of majorization \cite{Nielsen99,Jonathan-Plenio}.
Quantitatively, {DLC} accounts for the response of the  R\'{e}nyi entropy
\be
   S_\alpha \doteq { {1}\over{1-\alpha} }  \log \tr \rho_A^\alpha
\ee
to the changing of a control parameter $h$ in the Hamiltonian. Here, $\rho_A \doteq \tr_B |g\rangle \langle g|$ is the reduced density matrix of the block $A$ and $\alpha$ is a free parameter which tunes different entanglement measures \cite{NielsenChuang}. For instance, while low $\alpha$'s weight more evenly all eigenvalues of $\rho_A$, higher values of $\alpha$ enhance the role of the larger eigenvalues, which, {as} we are going to show, are more sensitive to the edge state behavior.

{DLC} holds if and only if all the $\alpha$-entropies are monotonous:
\be
   \partial_h S_\alpha \ge 0 \: , \qquad \forall \alpha\ge 0
   \label{localconvdef}
\ee or $\partial_h S_\alpha \leq 0$, $\forall \alpha \ge 0$ \cite{Turgut,Klimesh}.
{DLC} was first employed in a many-body problem by Cui {\it et al.} in \cite{Cui-Gu}.  An important motivation behind {\onlinecite{Cui-Gu}} is the observation that an adiabatic quantum algorithm \cite{adiabatic} may {exhibit greater computational capabilities only in a given phase that is not DLC}. Here, by identifying a physical mechanism responsible for lack of {DLC}, we argue that the computational power of a non-convertible phase is actually much bigger and that  a quantum phase cannot be exploited as an efficient (universal) quantum simulator  \cite{feynman1982} if it is locally convertible.

We  explain the phenomenology of local convertibility in quantum phases supporting edge states (that is, excitations localized at the boundaries of many-body system) \cite{hasan2010,mourik2012}.
We shall show that, in phases with boundary states, the operation of dividing the system into two partitions reveals long-ranged correlations that cannot be resolved within the partition. These are the manifestation of the edge states created at the boundaries between the sub-regions. These considerations are reflected by the non{-}trivial behavior of the entanglement entropy: for some (low) $\alpha$'s, the R\'{e}nyi entropies are sensitive to short-range entanglement and increase when a quantum phase transition (QPTs) is approached, but for other (large) $\alpha$'s, the entropies do the opposite. This implies that the entanglement between edge states can decrease, even as the correlation length increases.

\section{The quantum Ising chain}
To show the effect of edge states on local convertibility, it is desirable to have a model with three properties:
(i) it should support edge states; (ii) quasi-particle excitations should be clearly identifiable; and (iii) there should be a mechanism for destroying the edge states and observing the different behavior.
The one-dimensional transverse field Ising model fulfills these requirements  \cite{lieb1961,mcoy}.
It is defined by the Hamiltonian
\begin{eqnarray}
H_{\rm I}=-\sum_{j=1}^N \left( t \: \sigma_j^x \sigma_{j+1}^x + h \: \sigma_j^{z} \right) \; ,
\label{HIsing}
\end{eqnarray}
where $\sigma_j^\alpha$ are Pauli matrices, $t$ is a hopping amplitude (which we can set to $t=1$) and $h$ is the control parameter for the external magnetic field. A quantum phase transition for $h=t=1$ happens in the thermodynamic limit of $N \to \infty$.  This QPT's signatures have been recently observed experimentally \cite{coldea2010}.
We note that,  with an additional $\sigma _z\sigma _z$, and  by allowing a distribution of the couplings on a sparse graph, the Hamiltonian (\ref{HIsing}) would result  in a quantum Merlin-Arthur (QMA) -hard ground state problem, hence providing a universal quantum simulator \cite{Biamonte}. It has also been shown in \onlinecite{Biamonte} that, with all off-diagonal matrix elements in the standard basis being real and sharing the same sign, the model falls in the class of {so called}  stoquastic systems,
see Refs.\cite{Biamonte, Terhal} for related topics.

\begin{figure}[t]
\includegraphics[width=\linewidth]{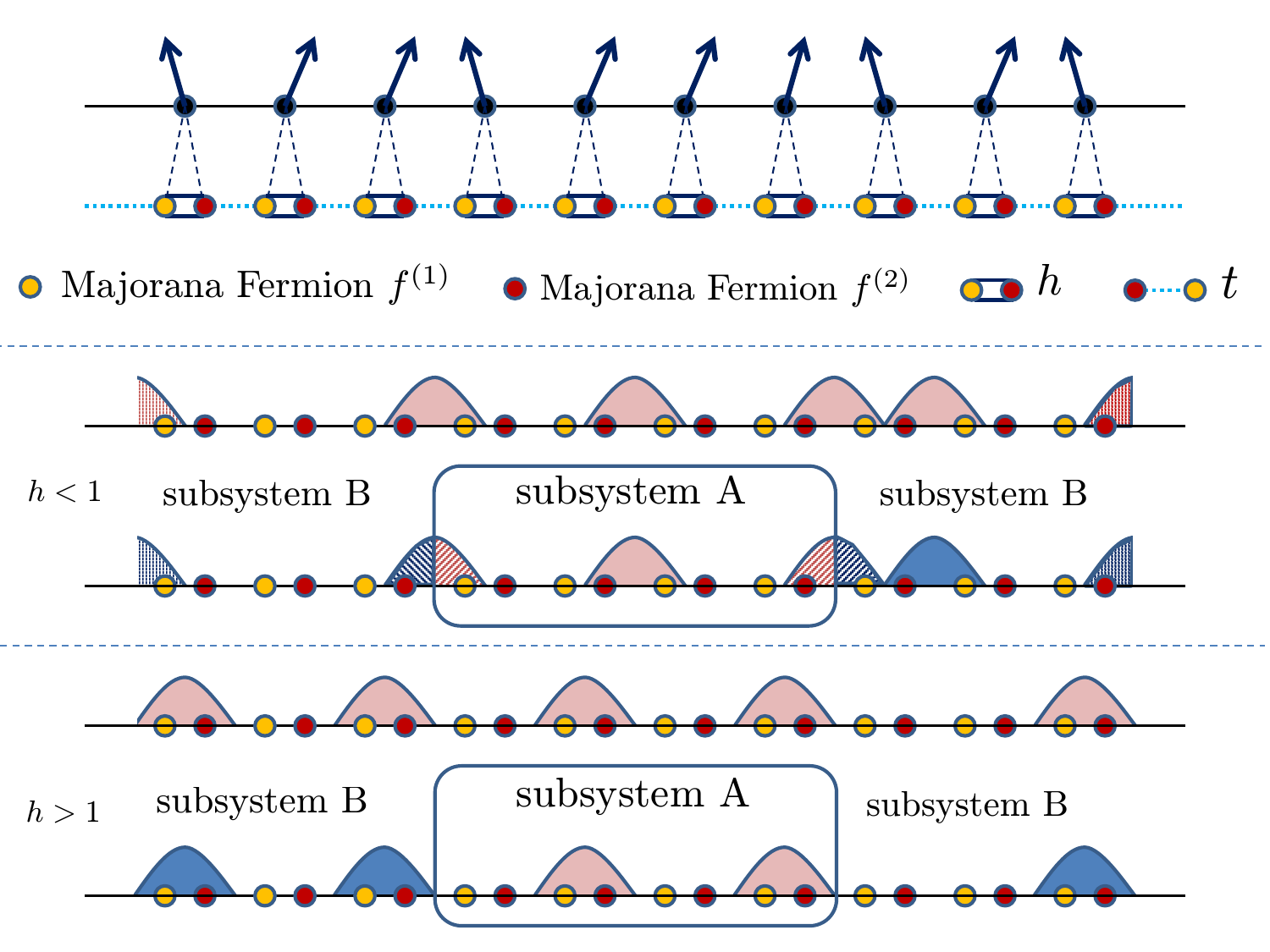}
\caption{\label{Majorana} Top: the Ising chain is mapped into a system of Majorana fermions by doubling the lattice sites. {Center} and bottom: a schematic of the  quasi-particle excitations in the two phases and the effect of bipartitioning the system; for small $h$,  edge states form at the opposite  boundaries of the subsystem A.
The property of local convertibility depends on the correlations between such edge states. }
\end{figure}

The Hilbert space acted on by Hamiltonian (\ref{HIsing}) can be described in terms of eigenstates of the string operator $\mu_N^x = \prod_{j=1}^N \sigma_j^z$, which generates the $\mathbb{Z}_2$ symmetry of (\ref{HIsing}). For $h>1$ the system is paramagnetic with $\langle \sigma^x \rangle =0$. For $h<1$, the spectrum of the Ising model becomes doubly degenerate. A ground state which is also an eigenstate of $\mu_N^x$ has a vanishing order parameter $\langle \sigma^x \rangle =0$. This ground state is   known as the `thermal ground state'. This is the {initial} state employed in the  2-SAT problem and in adiabatic quantum computation protocols for finite $N$ \cite{adiabatic}. In the thermodynamic limit ($N \to \infty$),  $\sigma^x$ can acquire a non-zero expectation value. The symmetry will be  broken spontaneously and the ground state will be  given by the (anti)symmetric combination of the two eigenstates of $\mu_N^x$. For $h<1$  we consider both the ferromagnetic ground state with non vanishing order parameter $\langle \sigma^x\rangle$ and the thermal one enjoying the same  $\mathbb{Z}_2$ symmetry as the Hamiltonian.

The quantum Ising model (\ref{HIsing}) can be mapped exactly, although non-locally, to  a system of free spinless fermions $\{c_j,c^\dagger_j\}$ \cite{lieb1961}. 
We remark that the mapping {in \cite{lieb1961}} preserves the entanglement between $A$ and $B$ \cite{vidal2003, latorre2004} and generates the Kitaev chain. As emphasized in \cite{kitaev2000}, this formulation highlights the presence of Majorana edge states as emergent degrees of freedom. Majorana fermions are the elusive particles (coinciding with their own anti-particles), proposed by E. Majorana. Many research groups are trying to find and manipulate them \cite{hasan2010, mourik2012}.
Each Dirac  fermion of the chain can be used to define two Majorana fermions:
\be
  f_j^{(1)} \equiv \left[ \prod_{l<j} \sigma_l^z \right] \sigma_j^x = c_j^\dag + c_j \, , \quad
  f_j^{(2)} \equiv \left[ \prod_{l<j} \sigma_l^z \right] \sigma_j^y =\ii \left( c_j^\dag - c_j \right) \;
  \label{majoranadef}
\ee

We represent this mapping pictorially in figure \ref{Majorana}. In the paramagnetic phase ($h>1$), the Hamiltonian pairs predominantly Majoranas on the same site $j$ (this correlation is drawn as a double line in the picture). In the ferromagnetic phase ($h<1$), the dashed line connecting different sites is dominant. In Kitaev's approach, the double degeneracy of this phase emerges as the first and last Majoranas are left unpaired  and can be combined into a complex fermion (the  occupancy/vacancy of this fermion  costs no energy). We will see that the same picture applies when the system is divided into two partitions: in the ferromagnetic phase this operation cuts the dominant link and leaves unpaired Majorana edge states on each side of the cut.

This is a key many-body feature that renders phases supporting boundary states more ``quantum'' than other systems. 
In fact, since any subsystem develops  edge states, in these phases qubits of information are stored non-locally between the sites and we will see that this is mirrored by the non-trivial entanglement behavior, yielding non-local convertibility. Such phenomenology must be a necessary ingredient of a machine aimed at simulating a generic quantum system and this is the reason for which {we believe} non-local convertibility to be a strong indicator of a higher computational power.

\section{The $\mathbb{Z}_2$ symmetric ground state}

\begin{figure}[t]
   \dimen0=\columnwidth
   \advance\dimen0 by -\columnsep
   \divide\dimen0 by 2
   \noindent\begin{minipage}[t]{\dimen0}
   \includegraphics[width=\columnwidth]{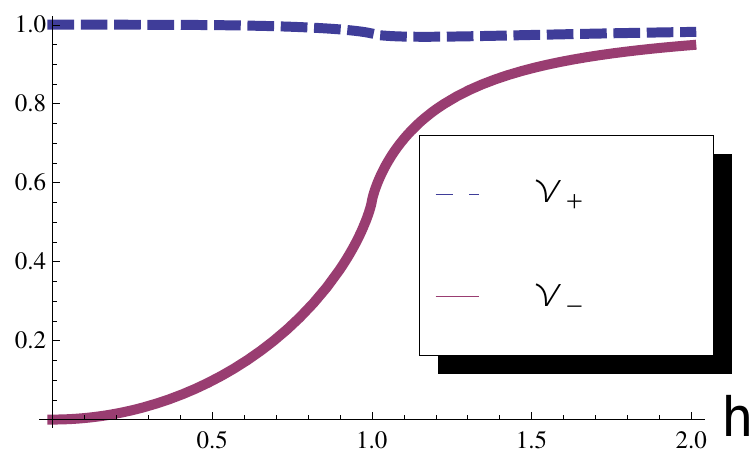}
   \end{minipage}
   \hfill
   \begin{minipage}[t]{\dimen0}
   \includegraphics[width=\columnwidth]{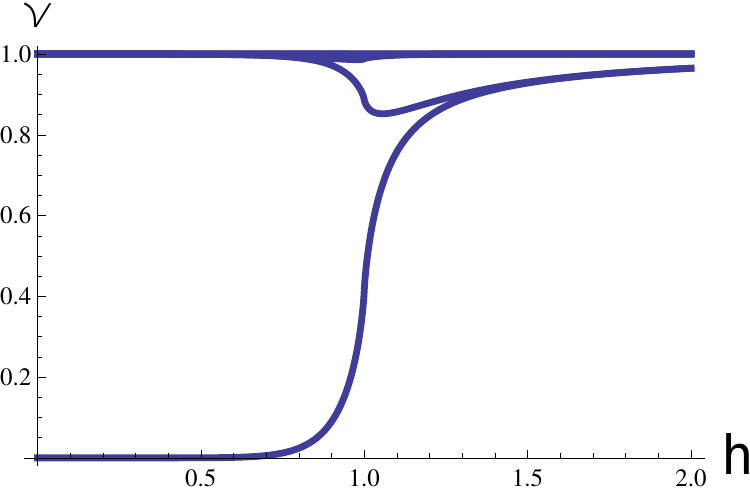}
   \end{minipage}
   \caption{\label{fig:correigen}Plot of the occupation number $\nu_j$ obtained from the correlation matrix (\ref{Bdef}) as a function of $h$ for $L=2$ (left) and $L=10$ (Right). For $L=2$, the explicit form of the eigenvalues $\nu_\pm$ is given in (\ref{nupm}). Notice that only one of the $\nu$'s exhibits non-trivial behavior: it corresponds to the boundary state, which is only partially contained in the subregion.}
\end{figure}

In figure \ref{fig:correigen} we plot these eigenvalues $\nu_j$ as a function of the magnetic field for As explained in the methods section, in the Ising chain, the $2^L$ states within a block of $L$ consecutive sites can be constructed in terms of individual quasi-particle excitations, which can be either occupied or empty.
These excitations are in general delocalized, with a typical size set by the correlation length. However, a  $\mathbb{Z}_2$ symmetric state possesses one special excitation, with support lying at the opposite edges of the block and formed by two Majorana edge states \cite{kitaev2000}. When the block is extended to the whole system ($L=N$), the block excitations coincide with the systems' excitations, including the boundary states.

The entanglement between two subsystems $A$ and $B$ can be extracted from the $2 L$ eigenvalues $\pm \ii \nu_j$ of the
correlation matrix Eq.~(\ref{Bdef}) incorporating the correlations of the excitations within the spin block. Here $L$ is the number of lattice sites in $A$. The eigenvalues of the reduced density matrix can then be constructed out of the $\nu_j$'s, using (\ref{lambdadef}) in the methods section. The $\nu$'s  can be interpreted as sort of occupation numbers, since they capture the overlaps between each block quasi-particle excitation and the ground state, according to (\ref{Bmodes}): $\nu_j =0$ means that this block excitation is half filled and half empty in the ground state, while $\nu_j = 1$ indicates that the excitation is either completely occupied or not present at all.

$L=2$ and $L=10$. Notice that in both cases only one block excitation has a non-trivial behavior, while the other eigenvalues stay approximately constant around unity in all phases.  Significant deviations happen only close to the QPT (as the correlation length diverges). As discussed, the modes with $\nu_j \simeq 1$ define bulk states.
In contrast, the non-trivial eigenvalue is close to zero for $h \simeq 0$ and rises rapidly toward $1$ crossing the QPT at $h=1$: in the ferromagnetic phase, it corresponds to an block excitation which is neither occupied, nor empty. By cutting the chain into two subregions, we severed the dominant inter-site correlation and hence generated two unpaired Majorana edge states (see Fig.~\ref{Majorana}). We noticed, however, that as $h$ increases, the occupation number of this edge excitation increase, indicating edge state recombination.

Having discussed the behavior of the eigenvalues $\nu_j$'s and the role of the boundary states, it is straightforward to analyze the R\'{e}nyi entropy and address the issue of differential  local convertibility.
It is interesting to concentrate on the two extreme limits: $L=2$ and $L \to \infty$.

\begin{figure}[t]
   \dimen0=\columnwidth
   \advance\dimen0 by -\columnsep
   \divide\dimen0 by 2
   \noindent\begin{minipage}[t]{\dimen0}
   \includegraphics[width=\columnwidth]{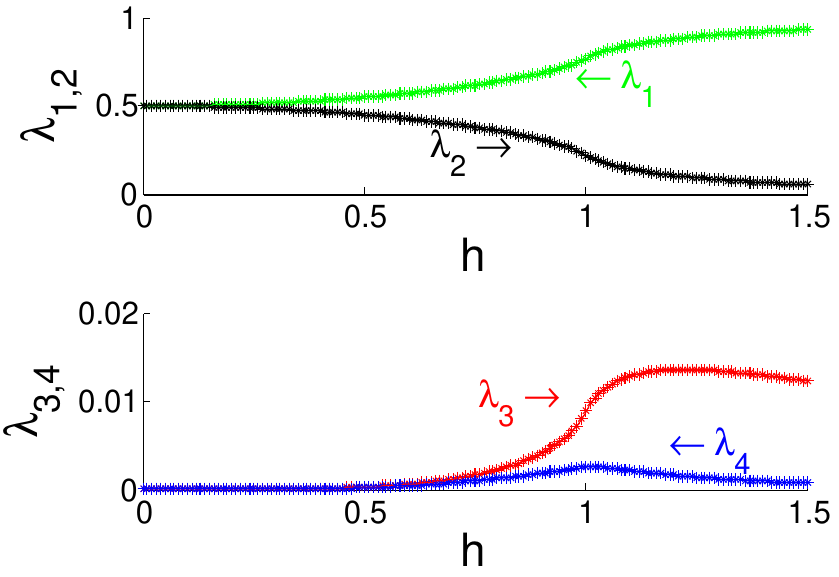}
   \end{minipage}
   \hfill
   \begin{minipage}[t]{\dimen0}
   \includegraphics[width=\columnwidth]{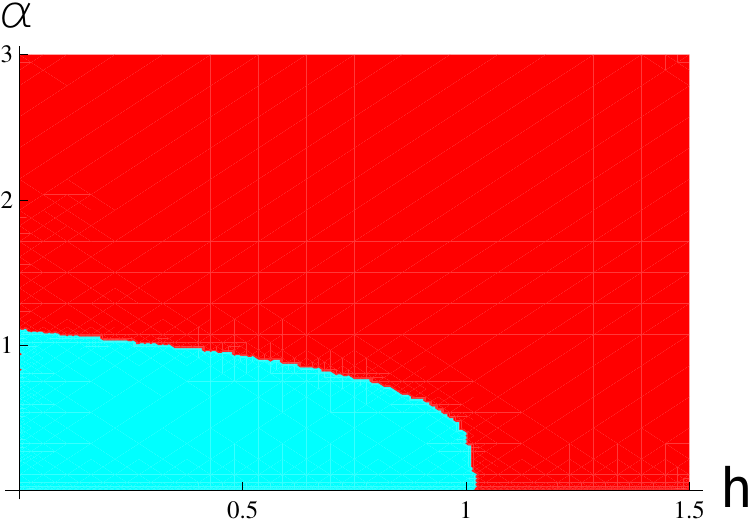}
   \end{minipage}
   \caption{\label{2spinAn} Left: Plot of the four eigenvalues of $\rho_A$ for $L=2$, as a function of $h$. The solid lines depict analytical results, while the crosses show the numerical results with N=200 (notice the different scales in the vertical axis between the top and bottom panels.). Right: Contour plot of the sign of the derivative with respect to $h$ of the R\'{e}nyi entropy for different values of $h$ and $\alpha$.}
\end{figure}

The two occupation numbers $\nu_\pm$ for $L=2$ are shown in the left panel of Fig.~\ref{fig:correigen} and the resulting four eigenvalues of the reduced density matrix, according to (\ref{lambdadef}), are plotted in the left panel of Fig.~\ref{2spinAn}. While in locally convertible phases the largest eigenvalue(s) decrease approaching the QPT, indicating an increase of the entanglement, here we see that the edge state recombination results in a growing larger eigenvalue. The right panel of  Fig.~\ref{2spinAn} presents the sign of the entanglement entropy derivative,
to be considered in relation with (\ref{localconvdef}).
We see that in the paramagnetic phase the R\'{e}nyi entropy always decreases. Instead, in the doubly degenerate phase the entropy derivative vanishes at some value of $\alpha$ and changes sign, indicating that local (differential) convertibility is lost in this phase (as already observed numerically for small $N$ and larger $L$ in \cite{Cui-Gu}).
It is important to notice here that these results imply that any operation acting on two sites alone effectively creates edge states at the boundary of the sites and hence projects the system onto these, loosing part of the coherence in the original state.

\begin{figure}
   \dimen0=\columnwidth
   \advance\dimen0 by -\columnsep
   \divide\dimen0 by 2
   \noindent\begin{minipage}[t]{\dimen0}
   \includegraphics[width=\columnwidth]{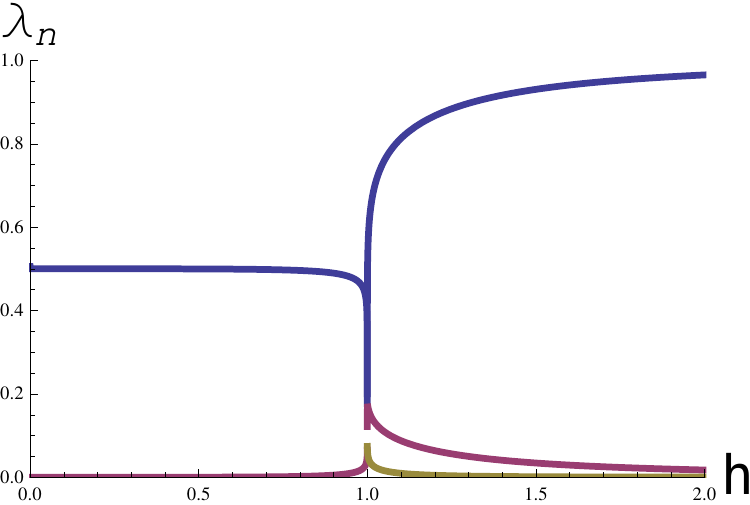}
   \end{minipage}
   \hfill
   \begin{minipage}[t]{\dimen0}
   \includegraphics[width=\columnwidth]{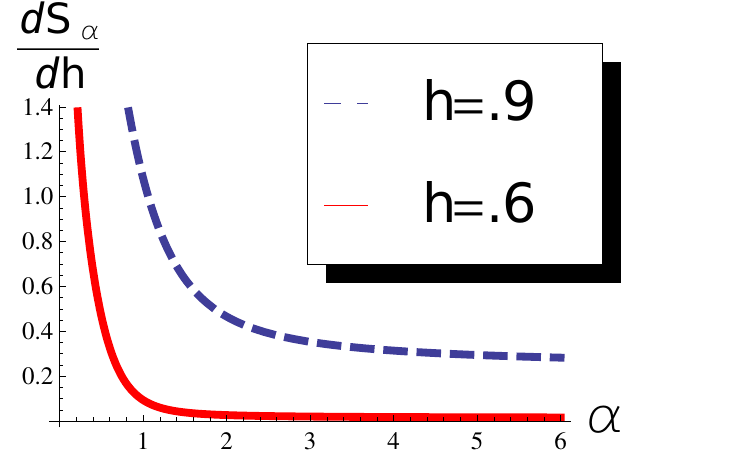}
   \end{minipage}
   \caption{\label{InfBlockAn}Left: Plot of the first few eigenvalues of $\rho_A$, for an infinite size block, as a function of $h$. The eigenvalues' multiplicities are not shown (for instance, the highest eigenvalue is doubly degenerate for $h<1$ and unique for $h>1$, see \cite{franchini10}) .  Right: Plot of the derivative of the R\'{e}nyi entropy with respect to the magnetic field $h$, as a function of $\alpha$, for two different values of $h$ in the ferromagnetic region.}
\end{figure}

For the $L \to \infty$ limit, we can take advantage of the results of \cite{its05, franchini08, franchini10}, where the full spectrum (eigenvalues and multiplicities) of the reduced density matrix and the R\'{e}nyi entropies were calculated analytically. Fig.~\ref{InfBlockAn} shows a plot of the first few eigenvalues of $\rho_A$ and a plot of the entropy derivative as a function of $\alpha$  for $h=0.6$ and $h=0.9$. We see that the largest eigenvalue (doubly degenerate in the ferromagnetic phase) decrease monotonously toward the QPT, while smaller eigenvalues are allowed to grow, yielding a monotonous increase of all the R\'{e}nyi entropies. It is thus clear that local convertibility is restored in the infinite $L$ limit.

We checked these results numerically for systems up to $N=200$ and with different partitions. We  considered different block sizes and  move the blocks within the chain. The qualitative picture does not change significantly as one varies $(A|B)$, but the location of the curve where the entropy derivative vanishes transitions in the $(h,\alpha)$ plane. It tends towards the phase transition line $h=1$ as the block sizes grow bigger, confirming our expectation on the role of the boundary excitations. Namely, we see that as long as the edge states from different boundaries do not overlap, their occupation number stays constant and vanishing.  It starts increasing only once the correlation length grows comparable to one of the block sizes,
indicating the recombination of the edge states and a decrease in the entanglement contribution from the edge states.

\section{Symmetry broken ground state}
\begin{figure}
   \dimen0=\columnwidth
   \advance\dimen0 by -\columnsep
   \divide\dimen0 by 2
   \noindent\begin{minipage}[t]{\dimen0}
   \includegraphics[width=\columnwidth]{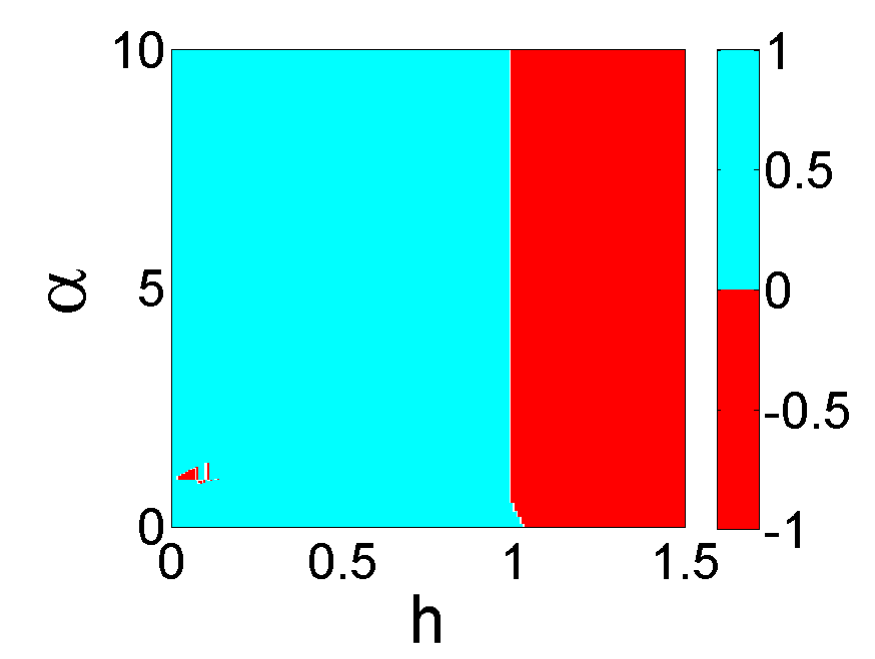}
   \end{minipage}
   \hfill
   \begin{minipage}[t]{\dimen0}
   \includegraphics[width=\columnwidth]{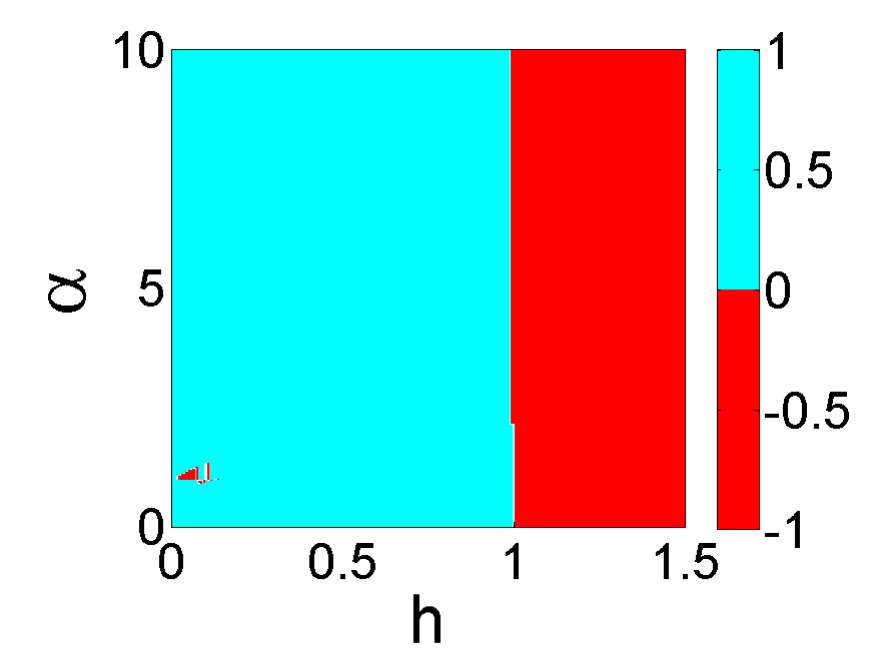}
   \end{minipage}
\caption{\label{symmetrybroken}(Color online) Numeric results of differential local convertibility for the ferromagnetic (symmetry broken).
\blue{Left:} a partition $200=2|198$, the right $200=50|100|50$.}
\end{figure}

To further confirm our interpretation on the role of boundary modes, in the ordered phase $h<1$ we also considered the ferromagnetic ground state for which $\langle \sigma^x \rangle \neq 0$. Since this state does not support well defined Majorana edge states, we expect a restoration of local convertibility.
We  calculate the R'{e}nyi entropy of this symmetry broken ground state numerically. Namely we add a very small perturbation $\epsilon(\sigma_1^x+\sigma_N^x)$ to the Hamiltonian (\ref{HIsing}) and apply the variational matrix product state routine to obtain the ground state \cite{mps2008}. In this work the converge tolerance is $10^{-6}$.
Fig.~\ref{symmetrybroken} shows the plots of the sign of the entropy derivative for two possible partitions (small and large $A$ block) and {this} validates our expectation that both phases are locally convertible. We considered several partitioning choices and the results are not distinguishable from those in Fig.~\ref{symmetrybroken}.

In conclusion, we see that for $h>1$ the disordered ground state is always locally convertible. In the ordered phase, the ferromagnetic ground state, i.e. with broken symmetry, is also locally convertible for any chosen partition. For the thermal ground state, however, the convertibility depends on the interplay between the size of the partitions $(A|B)$  and the correlation length of the system. This phenomenon is a manifestation of edge state recombination.
These entangled pairs lie on opposite boundaries of the partition (see Fig.~\ref{Majorana}), but with a finite support intruding in the bulk about the order of the correlation length.
For sufficiently large block size, the entanglement between boundary states does not depend on the correlation length and remains constant throughout the phase.
However,  as this length increases approaching a QPT, the edge states effectively grow closer.  If either of the subregions $A$ or $B$ is sufficiently small, the tails of these states can overlap and we see their occupation number increasing and their entanglement decreasing, yielding non-local convertibility.

\section{Conclusions and discussions}
We considered a specific notion of long range entanglement, defined as {the lack of} differential local convertibility for a bipartite system {$(B|A|B)$}.
In phases where differential local convertibility holds,  the response of the ground state to an external perturbation can be rendered by local means (in $A$ and $B$), and hence such phases offer {more restricted computational capability under adiabatic perturbation}.
By identifying a mechanism that breaks {differential} local convertibility, we discussed the role of edge states in determining quantitatively whether the adiabatic perturbation of a quantum many-body system can supersede a protocol limited to LOCC operations between $A$ and $B$.

In the class of models we consider, we found that destroying edge states (for instance, by breaking the symmetry of the Hamiltonian) yields locally convertible quantum phases. In contrast, in quantum phases with edge states, the convertibility depends on the interplay between the sizes of the partitions, and the correlation length of the system.  Approaching the QPT, at some point the correlation length grows to be comparable with one of the block sizes and the boundary states start recombining: from this point forward, their contribution to the entanglement start decreasing. For sufficiently large $\alpha$'s, the R\'{e}nyi entropies are dominated by the edge state behavior and they decrease toward a QPT. At the same time, low-$\alpha$ entropies still increase as usual, because more states are required to construct the reduced density matrix in the Schmidt decomposition \cite{NielsenChuang}. This combined behavior leads to a breakdown of local convertibility.
Summarizing: the entanglement of bulk states increases toward a QPT, while that of edge states does the opposite.
Since classical manipulations can never increase entanglement, this contrasting behavior signals the existence of genuine quantum character in these states.
We remark that in LOCCs ``local'' means that the manipulations are restricted within each region of the bipartition, and thus can still involve entangling operations, while in physical applications, locality has a much stricter meaning. For this reason, we believe that our observation that the edge states recombination scheme works already for the smallest partitions $L=2$ (Fig. \ref{2spinAn}), has an important implication: the construction of quantum circuits already in terms of two-qubit gates generates a long-range coherence (between the Majorana fermions at the edges of the block) that, if incorporated,  may  provide an added value for the computation in the quantum phase.  The two quantum phases possesses qualitatively different computational capabilities: indeed, it was shown in \cite{Chen} that to connect them one needs a quantum circuit whose depth scales at least linearly with the system's size; in contrast, different ground states within the same quantum phase can be connected with a depth sub-linear in $N$ \cite{Chen, Wen}. As a step forward in this scenario,  we identify in the local convertibility the `bit' providing the qualitative difference  in resources between the two quantum phases.

Because of the fundamental relation between quantum circuits and Hamiltonian satisfiability problems\cite{Kitaev_QMA,Aharonov},  we believe that our results  are relevant in the framework of the theory of quantum complexity classes\cite{kempe, bravy, gu, Qcomplexity}. Indeed, universal quantum computers encompass Hamiltonian models within the Quantum-Merlin-Arthur (QMA) complete complexity class. Notwithstanding,  the quantum Ising model we considered in the present article is believed to be QMA-complete only if it is suitably generalized \cite{gottesman,Biamonte,Terhal},
our results indicate that the edge states should survive in these more general models and their convertibility properties will be investigated in the future. It is also interesting to understand whether these more general models can be reached adiabatically starting from the quantum Ising chain, without crossing a QPT.

Incidentally, we comment that a similar lack of {differential} local convertibility has been observed recently in topologically ordered systems \cite{Cui-Amico, Hamma-Cincio}. In Ref. \onlinecite{Cui-Amico},  in particular, it was demonstrated how the phase with non-trivial edge states, the  Haldane phase, is indeed non locally convertible; given the role of the Haldane order in systems of cold atoms with dipolar interaction, this opens the avenue towards experimental verifications of the convertibility protocols. Indeed, topological systems are characterized by a form of long range entanglement, as a finite depth quantum circuit can disentangle the system only at short-range \cite{Wen, Hastings}. Compared to this, we remark that local convertibility seems to detect a form of long-range entanglement, even while working on relatively small systems.
We believe further research is called for, to understand the exact relation between local convertibility and the form of long-range entanglement, as defined in \cite{Wen}.

To conclude, the universal quantum simulator, as envisioned by Feynman\cite{feynman1982},  was thought for  simulating all quantum interactions, hence superseding in efficiency any classical algorithm.
Our theory shows that edge states provide phases that are not locally convertible and this indicates that a quantum simulator should show similar non-convertibility. In this respect, it is not accidental that protected edge states  play a crucial role in quantum computation (although it is clear that the existence of edge states is only a necessary condition for a system to be viable as a universal quantum simulator)  \cite{nayak2008}.
This should direct further efforts toward the identification of workable quantum simulators and could open a way for a next generation, specifically designed quantum algorithms.

\begin{acknowledgments}
We thank Alioscia Hamma for several fruitful discussions. F.F. thanks Chen Fang for his interesting seminar and his insights. F.F. was supported by a Marie Curie International Outgoing Fellowship within the 7th European Community Framework Programme (FP7/2007-2013) under the grant PIOF-PHY-276093. J.C. acknowledges  {Max-Planck-Gesellschaft Rechenzentrum} Garching for the computational resource{, and thanks Mari Carmen Ba\~nuls for discussions on numerical simulation.}  H.F. is supported by 973 program (2010CB922904) and grants from NSFC and CAS. V.K. is supported by the NSF DMS-1205422. L.C.K. and V.V. are supported by the National Research Foundation \& Ministry of Education, Singapore. M.G. is supported by
National Basic Research Program of China Grant 2011CBA00300, 2011CBA00302, the National Natural Science Foundation of China Grant 11450110058, 61033001, 61061130540.
L.C.K. and L.A. are also supported by the National Institute of Education and the Institute of Advanced Studies, Nanyang Technological University, 1 Nanyang Walk, Singapore 637616.
\end{acknowledgments}

\appendix

\section{Methods}

\subsection{The R\'{e}nyi entropies}

An advantage of  working with a quadratic theory such as the Ising chain is that many-body states can be constructed exactly out of individual quasi-particle excitations. The latter can be found as the linear combination of the fermionic operators $\{ c_j , c_j^\dag \}$ in (\ref{majoranadef}), which diagonalize the Hamiltonian. Doing so, we define a new set of operators $\{ \tilde{c}_j , \tilde{c}_j^\dag \}$ so that  the ground state $|g\rangle$ is  annihilated by all $\tilde{c}_j$. Additionally, one can excite quasi-particles by progressively applying all possible combinations of $\tilde{c}_j^\dag$, yielding a total of $2^N$ states in the Hilbert space.

To calculate the entanglement between the subregions $A$ and $B$, we {employ} the Schmidt decomposition of the ground state
\be
   | g \rangle = \sum_l \sqrt{\lambda_l} \, | \psi_l^{(A)} \rangle \, | \psi_l^{(B)} \rangle \; ,
\ee
where $|\psi_l^{(A,B)} \rangle$ span the Hilbert space of block $A$ and $B$ respectively \cite{NielsenChuang}. We are after the eigenvalues $\lambda_l$, which can be found, for instance, as
\be
   \lambda_l = \langle g | \psi_l^{(A)} \rangle \langle \psi_l^{(A)} | g \rangle \; ,
   \label{lambdaproj}
\ee
where a trace over the $B$ degrees of freedom is implicit.
Similarly to what is done for the entire system, the states $|\psi_l^{(A)}\rangle$ can be constructed in terms of individual excitations. However, these differ from those of the whole chain, as they are contained inside the block. If $A$ consists of $L$ consecutive sites, these block excitations $\{ d_j , d_j^\dag \}$ are the linear combinations of the $c$-operators within the block, that diagonalize the correlation matrix constructed out of all their two-point correlation functions, as shown below. Each state $|\psi_l^{(A)}\rangle$ of this $2^L$-dimensional Hilbert space can thus be characterized by the occupation number $0$ or $1$ of each block excitation. Moreover, the eigenvalues $\nu_j$ of the aforementioned correlation matrix provide us with the expectation values
\be
   \langle g| d_j d_j^\dagger | g \rangle = {1 + \nu_j \over 2} \; , \qquad
   \langle g| d_j^\dagger d_j | g \rangle = {1 - \nu_j \over 2} \; ,
   \label{Bmodes}
\ee
with all other correlations being zero. Note that $\nu_j \simeq 1$ indicates that $d_j$ annihilates the vacuum $|g\rangle$. It follows that certain quasi-particle excitations of the Hamiltonian are completely contained within the block, since, $d_j | g\rangle = 0 $ implies that $d_j$ is just a superposition of $\tilde{c}_j$'s. Since $d_j$ is defined just within the block, it follows that these $\tilde{c}_j$'s are also contained in the block. Conversely, smaller values of $\nu_j$ are related to excitations lying only partially within a subregion.
In turn, $d_j d_j^\dag$ acts on the ground state as a projection operator which selects the component with $0$ occupation number for the $l$-th block excitation, while $d_j^\dag d_j$ projects it onto an occupied $l$-th excitation. Hence, (\ref{lambdaproj}) can be written as the expectation value of a string of operators of this type. Using (\ref{Bmodes}) as the building blocks of these correlators, we have
\be
   \{ \lambda_l \} = \prod_{j=1}^L \left( {1 \pm \nu_j \over 2} \right) \, ,
   \label{lambdadef}
\ee
with all the possible combinations of plus/minus signs, corresponding to the occupation/unoccupation of the different block excitations.

Finally, the R'{e}nyi entropies read  \cite{vidal2003, latorre2004}
\be
   S_\alpha (\rho_A) = {1 \over 1- \alpha} \sum_{j=1}^L \log \left[
   \left(  {1 + \nu_j \over 2} \right)^\alpha
   + \left(  {1 - \nu_j \over 2} \right)^\alpha \right] \; .
   \label{Salpha}
\ee

\subsection{The Correlation Matrix}

The R\'{e}nyi entropies are accessed through the eigenvalues of the reduced density matrix of a block of $L$ consecutive spins for the thermal ground state \cite{vidal2003, latorre2004}. Such eigenvalues can be obtained from the diagonalization of the $2L \times 2L$ correlation matrix: $ \langle f_k^{(a)} f_j^{(b)} \rangle =\delta_{j,k} \delta_{a,b} + i \left( {\cal B}_L \right)_{(j,k)}^{(a,b)}$, with
\be
   {\cal B}_L \equiv \left( \begin{array}{cccc}
                              \Pi_0 & \Pi_1 & \ldots & \Pi_{L-1} \cr
                               \Pi_{-1} & \Pi_0 && \vdots \cr
                               \vdots & & \ddots & \vdots \cr
                               \Pi_{1-L} & \ldots & \ldots & \Pi_0 \cr
                               \end{array} \right),
  \label{Bdef}
\ee
where $j,k$ specifies the entry $\Pi_{j-k} \equiv \left( \begin{array}{cc} 0 & g_{j-k} \cr - g_{k-j} & 0 \cr \end{array} \right)$, which is itself a $2 \times 2$ matrix whose $a,b$ entries are defined as
\be
    g_j \equiv {1 \over 2 \pi} \int_0^{2 \pi} { \cos \theta -h  + \ii \sin \theta \over
    \sqrt{ (\cos \theta - h )^2 +  \sin^2 \theta} } \: \eu^{\ii j \theta} \: \de \theta \: .
   \label{gjdef}
\ee

The anti-symmetric matrix ${\cal B}$ can be brought into a block-diagonal form by a $SO(2L)$ rotation, with each block of the form $$\tilde{\Pi}_j = \nu_j \left( \begin{array}{cc} 0 & 1 \cr - 1 & 0 \cr \end{array} \right)$$ This rotation defines a new set of Majorana fermions $\tilde{f}_j^{(a)}$ with only pair-wise correlations. This rotated operator basis  can be used to introduce a new set of complex operators: $d_j = \left( \tilde{f}_j^{(1)} + \ii \tilde{f}_j^{(2)} \right)/ 2$ (and its hermitian  conjugate). The matrix (\ref{Bdef}) contains all information to completely solve the model. By taking $L =N$, i.e. extending the correlation matrix to the whole system, the $d$-modes coincide with the $\tilde{c}$-operators one would obtain from the diagonalization of the Hamiltonian (\ref{HIsing}).

For $L=2$, the two eigenvalues of the correlation matrix are easily found to be
\be
  \nu_\pm = \sqrt{ \left({g_1 - g_{-1} \over 2} \right)^2 + g_0^2}  \pm {g_1 + g_{-1} \over 2} \; ,
   \label{nupm}
\ee
which allows for a complete analytical study of the entanglement entropy and its derivative (see Fig. \ref{2spinAn}).

\subsection{Local convertibility and R\'{e}nyi entropies.}
\begin{figure}
   \dimen0=\columnwidth
   \advance\dimen0 by -\columnsep
   \divide\dimen0 by 2
   \noindent\begin{minipage}[t]{\dimen0}
   \includegraphics[width=\columnwidth]{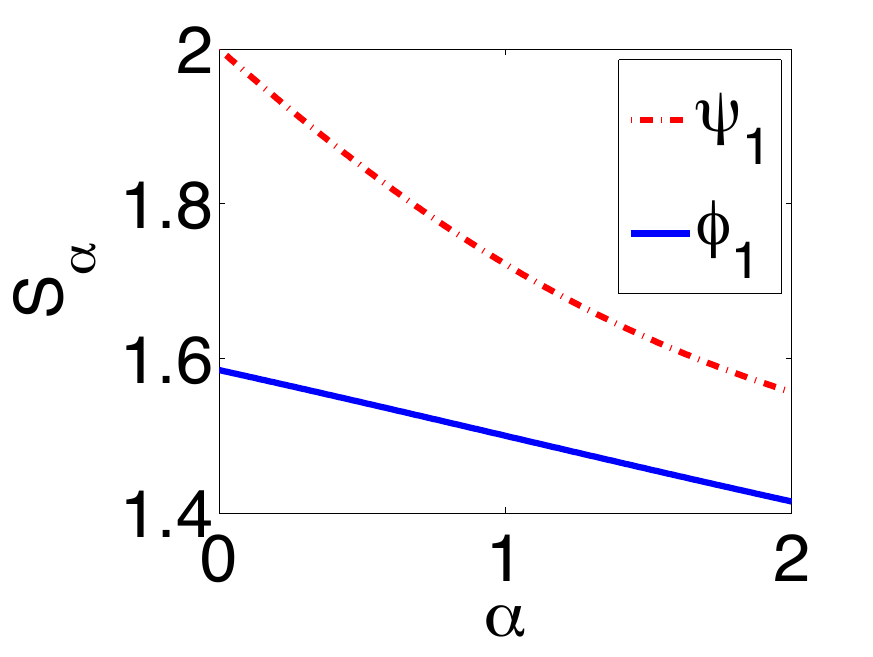}
   \end{minipage}
   \hfill
   \begin{minipage}[t]{\dimen0}
   \includegraphics[width=\columnwidth]{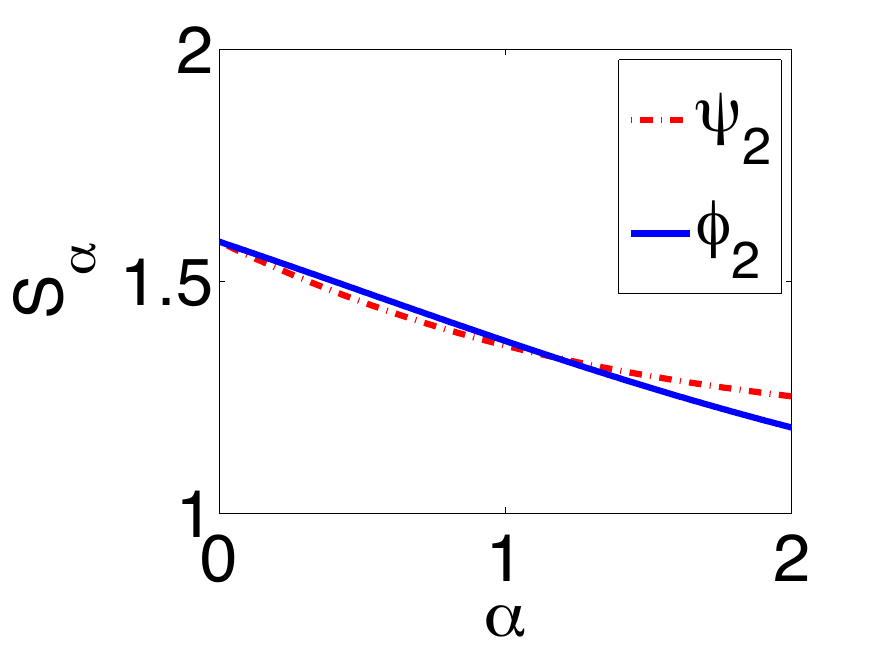}
   \end{minipage}
\caption{\label{schematic} The crossing or non-crossing of entanglement R\'{e}nyi entropies of two bipartite states $|\psi \rangle$ and
$|\phi \rangle$ may demonstrate two types of behavior{s}:  if the entropy curves do not touch, $|\psi _{1}\rangle $ can be locally converted to $|\phi_{1}\rangle $, see (\ref{states1}); if the two curves intersect (right), the two states, like those in (\ref{states2}),
cannot be locally converted to each other.}
\end{figure}

As we mentioned, one key property of entanglement measures is that they do not increase under local quantum operations and classical communication (LOCC). Considering pure bipartite states, for instance
\bea
   |\psi _1\rangle &  = &\sqrt {0.4} \, |11\rangle + \sqrt {0.4} \, |22\rangle +\sqrt {0.1} \, |33\rangle +\sqrt {0.1} \, |44\rangle \: , \nonumber \\
   |\phi _1\rangle & = & \sqrt {0.5} \, |11\rangle + \sqrt {0.25} \, |22\rangle +\sqrt {0.25} \, |33\rangle \: .
   \label{states1}
\eea
While it is not possible to convert $| \phi_1 \rangle$ into $| \psi_1 \rangle$ by LOCC, it is possible to operate locally on $|\psi_1 \rangle$ to transform it into $| \phi_1 \rangle$  (if assisted by the ancillary entangled state $|a\rangle=0.6|00\rangle+0.4|11\rangle$)  \cite{Jonathan-Plenio}.
If instead we take the two states
\bea
   |\psi _2\rangle & = & \sqrt {0.5} \, |11\rangle + \sqrt {0.4} \, |22\rangle +\sqrt {0.1} \, |33\rangle \, , \nonumber  \\
   |\phi _2\rangle & = & \sqrt {0.6} \, |11\rangle + \sqrt {0.2} \, |22\rangle +\sqrt {0.2} \, |33\rangle \, ,
   \label{states2}
\eea
it can be shown that they cannot to be interconverted by LOCC \cite{Nielsen99}, even with ancillas.
These properties can be captured  through the entanglement R\'{e}nyi entropies.
The necessary and sufficient condition for $|\phi \rangle$ to be locally convertible to $| \psi \rangle$ is that the inequality holds  \cite{Turgut}
\be
   S_{\alpha }([\psi ]) \ge S_{\alpha }([\phi ]) \quad {\rm for} \; {\rm all  } \; \alpha >0 \; .
   \label{major}
\ee
In this case, state $|\psi _{AB}\rangle $ can be converted locally to another
state $|\phi _{AB}\rangle $ of lower entanglement, possibly assisted by a ``catalyst'' state.
The above examples can be understood clearly by looking at their entanglement R\'{e}nyi entropies, as show
in Fig.~\ref{schematic}.

The notion of differential local convertibility was first discussed  in a many-body scenario in Ref. \cite{Cui-Gu}. Accordingly,  let   $|GS(g)\rangle$ be the ground state  of a given Hamiltonian $H(g)$, depending on a control parameter $g$. As $g$ is varied, it is questioned whether the adiabatic evolution of  $|GS(g)\rangle$ into the ground state of $H(g+\delta)$,  $|GS(g+\delta)\rangle$, can be simulated through assisted LOCC in $A$ and $B$ (bi-partitioning the system as explained in the introduction).  As explained above, the answer to this question lies in the majorization condition (\ref{major}) on the R'{e}nyi entropies, which is equivalent to
\be
  \partial_g S_{\alpha}(g)\leq 0 \qquad  {\rm (or} \; \partial_g S_{\alpha}\geq 0 {\rm )} \; , \qquad
  \forall \alpha \ge 0 \; .
\ee
This property is thus called differential convertibility.
Besides its own interest,  it is clear that this question parallels the protocol of an adiabatic quantum computation \cite{adiabatic}, since both consider the evolution of the instantaneous ground state as the control parameter is varied. While the interplay between local convertibility and other computational protocols remains to be understood, our analysis establishes a criterion to distinguish quantum phases where the adiabatic protocol is a particularly powerful tool.

\end{document}